# Phishing Dynamic Evolving Neural Fuzzy Framework for Online Detection "Zero-day" Phishing Email

[1]Ammar ALmomani;  [3]B. B. Gupta;  [1,2]Tat-Chee Wan;  [1]Altyeb Altaher;  [1]Selvakumar Manickam

[1]National Advanced IPv6 Centre (NAV6), Universiti Sains Malaysia, 11800 USM, Penang, Malaysia
[2]School of Computer Sciences, Universiti Sains Malaysia, 11800 USM, Penang, Malaysia
[3]Department of Computer Science, University of New Brunswick, Canada
Ammarali, tcwan, altyeb, selva@nav6.usm.my;  [3]gupta.brij@gmail.com

## Abstract

Phishing is a kind of attack in which criminals use spoofed emails and fraudulent web sites to trick financial organization and customers. Criminals try to lure online users by convincing them to reveal the username, passwords, credit card number and updating account information or fill billing information. One of the main problems of phishing email detection is the unknown "zero-day" phishing attack, (we define zero-day attacks as attacks that phisher mount using hosts that do not appear in blacklists and not trained on the old data sample and it is a noise data), which increases the level of difficulty to detect phishing email. Nowadays, phishers are creating different representation techniques to create unknown "zero-day" phishing email to breach the defenses of those detectors. Our proposed is a novel framework called phishing dynamic evolving neural fuzzy framework (PDENF), which adapts the evolving connectionist system (ECoS) based on a hybrid (supervised/unsupervised) learning approach. PDENF adaptive online is enhanced by offline learning to detect dynamically the phishing email included unknown zero-day phishing e-mails before it get to user account. PDENF is suggested to work for high-speed "life-long" learning with low memory footprint and minimizes the complexity of the rule base and configuration with few number of rules creation for email classification. We expect to achieves high performance, including high level of true positive, true negative, sensitivity, precision, F-measure and overall accuracy compared with other approaches.

**Keywords**: Phishing email, detection, zero-day, evolving connectionist System (Ecos).

## 1. Introduction

Email has been an online 'killer application' utilized by people, businesses, Governments and different organizations for the needs of communicating, sharing and distributing data (MAAWG, 2011). Phishing email is a subset of spam which is related to social engineering schemes, which depends on forged e-mails (i.e. claims that originated from a legitimate company or bank) and then through an embedded link within the e-mail, the phisher tries to redirect users to fake Websites. These fake Web sites are designed to obtain financial data from their victim fraudulently, including usernames, passwords, and credit card numbers, Occasionally, the phisher tries to misdirect the user to a fake website or to a legitimate one monitored by proxies (APWG, 2010).

The problem of phishing e-mail is becoming worse. A survey by (GARTNER, 2007) on phishing attack showed that about 3.6 million users in the USA lose money because of phishing. The total losses amount is estimated at US$3.2 billion dollars and the number of victims increased from 2.3 million in 2006 to 3.6 million in 2007. One of the newest reports depends on the eCrime Trends Report (Reports, 2012), which explain that Phishing attacks has increased by 12% year-over-year. Many problems arise due to phishing e-mails, most of which affect financial companies and their clients. Phishing is capable of damaging electronic commerce because it causes users to lose their trust on the Internet.

Today, one of the main problems in e-mails is the so-called unknown "zero-day" phishing e-mails**.** Zero-day attacks are defined as attacks that phishers mount using hosts that are not blacklisted or using techniques that evade known approaches in phishing detection (Bimal Parmar, 2012; Cook, Gurbani, & Daniluk, 2009; Dunlop, Groat, & Shelly, 2010; Khonji, Iraqi, & Jones, 2012a). Phishing e-mail is so complex that it cannot be detected by many of current techniques because the phisher can use new vulnerabilities which are never seen before(US-CERT, 2012). There are a number of possible solution to phishing but not effective yet(Venkatesh Ramanathan, 2012). These ranges from communication-oriented approaches like authentication protocols over blacklisting to content-based filtering approaches which usually depend on some of Artificial Intelligence (AI) techniques (Bergholz et al., 2010). Current AI algorithms are able to detect phishing e-mail based on fixed features and rules while a few number of machine learning algorithms design to work in online mode (N. Kasabov, 2005). The level of errors in the classification process will increase over time, especially when dealing with unknown zero-day phishing e-mails.

Phishing e-mail detection has been a major area of focus in a number of studies. In this proposed a framework called Phishing Dynamic Evolving Neural Fuzzy Framework (*PDENFF*) is proposed, where a dynamic process is one that continuously changes and progresses. This Framework is capable of determining dynamically whether e-mail is phishing or ham. The implementation of the proposed framework adapts the evolving clustering method (ECM) as a part of the dynamic evolving neural fuzzy inference system (DENFIS) in an online mode(N. Kasabov & Song, 2002; snjezana soltic, 2006) along the dynamic neural fuzzy inference system (DyNFIS) to enhance the rule creation in an offline mode (Y. C. Hwang & Q. Song, 2009). The proposed framework can detect phishing email by evolving stream data mining that leads to improve classification performance. It has a high level of performance and characterized by life-long learning with low memory footprint.

## 2. Related works

Phishing emails filtering methods depends on classification techniques which can be managed by several ways, such as features extraction, machine learning technique and clustering methods. for detecting phishing emails many approaches have been proposed, features extraction technique supposed by (Fette, Sadeh et al. 2007), his approach called (PILFER) method correlated with machine learning technique depend on features extraction to distinguish the phishing emails from ham(legitimate) emails . Fette used 10 features represent the phishing email features, then by using a random forest as a classifier to create a number of decision trees, PILFER able to detect the type of new email have the same style of features. The accuracy in this model has more than 96%, with false positive rate 0.1% and 4% false negative rate, but this technique still weak to detect zero-day phishing email because it depend on supervised learning algorithm.

Resent researcher depends on machine learning technique for detecting phishing emails. There are three types of machine learning technique used in field of phishing email included supervised learning, unsupervised learning and some of them used hybrid learning based on classifiers. The main rule of the classifiers depends on learning several inputs or features to expect a desirable output.

(Abu-Nimeh, Nappa et al. 2007) compared six classifiers related with machine learning technique for phishing email prediction the result of his studying denoted that there is no standard classifiers for phishing prediction. Another widely deployed technique used multi classifier related with machine learning for phishing email detection is (Saberi, Vahidi et al. 2007), the proposed method accuracy detected 94.4% of phishing emails. An another approach depend on three tier classification to detect phishing emails is Islam(Islam, Abawajy et al. 2009), if the first two if the first two classifier can't classify well the final tier will have the final decision, the average accuracy of this approach reach up to 97%, however, this approach consuming time and memory .

Other approaches used clustering method in phishing email detection method by (Dazeley, Yearwood et al. 2010). His proposed depends on shared method between unsupervised clustering algorithms with supervised classification algorithms then train the data by consensus clustering,. This technique increased the speed of classification with better accuracy than k-means algorithm. However, k-means algorithm design to work in offline mode, but it cannot work in online mode.

## 3. Problem statement

Compared to spam filtering, a few researches have been done so far in detecting phishing emails, which distinguish them from ham emails. One of the most critical aspects to distinguish between phishing and ham email is unknown "zero-day" phishing email before it get to user, because the phisher is able to use unknown features or techniques in his/her attack. The current approaches have many problems to deal with phishing email included unknown "zero-day" attack, which causes high level of false positives (*FPs*), false negatives (*FNs*) and low level of accuracy in classification process. However, *FP* denotes non-phishing e-mails marked as phishing, whereas *FN* represents the misidentification of a phishing e-mail.

There is other problems related with cost (i.e. *FP* is more expensive than *FN)*.In addition to this, there are problems of time consumption in classification process, complexity and the huge number of rule created from the most machine learning technique. There are other problems related with consuming memory or storage, especially in learning process as a life-long working to detect a new phishing emails. For example Artificial neural networks (*ANN*) techniques suffer from difficulties with selecting the structure of the phishing email or forgetting previously learned knowledge after further training. This overhead problem could be reduced using PDENFF that is proposed in this proposed.

## 4. Research motivation

Detection of zero-day attacks that were not caught by existing filters can't be determined in an easy way. This motivated the researcher to build a new methodology which can be able to detect the unknown "zero-day" phishing emails in online mode.

More motivation included:

1. The increase in interest in adaptive auto learning approaches as an efficient technology in Internet security and monitoring fields, which can be implemented to distinguish between phishing email and ham email in online mode and high speed.

2. The need to enhance the level of accuracy and trust for financial organizations by decreasing the level of phishing email attack.

3. The choice of a suitable online framework which can be able to work in real world and life-long working with low memory footprint and trying to put solutions for problems ofother techniques.

## 5. Objectives

The main goal of this proposed is to build a framework for evolving stream data mining that leads to improve classification performance and various capabilities. The focus is to apply this framework to solve the problem of phishing email, especially to detect the unknown "zero-day" of phishing emails in online mode, with the following objectives:

1. To develop a new on-line based learning method with low computational cost and to analyses its performance under different scenarios.
2. To improve the performance and accuracy in terms of the classification and prediction of phishing e-mail in the future.
3. To optimize memory consuming in the classifier process and to reduce the time needed for classify the email with unlimited learning, while the characteristics of phishing e-mail features have changed. This includes minimize the complexity of the rule-base and configuration.
4. To evaluate the proposed framework against using approaches for the purpose of phishing email detection.

## 6. Contribution

The main expected contribution of this proposal is a novel framework called phishing dynamic evolving neural fuzzy framework (PDENF), which adapts the evolving connectionist system (ECoS) (Kasabov 2007) based on a hybrid (supervised/unsupervised) learning approach. This framework has many sub-contributions in field of phishing email detection as follow.

1. Adaptive online is enhanced by offline learning to build lifelong learning system able to detect dynamically the unknown zero-day phishing e-mails without prior knowledge of the phishing e-mail itself.
2. propose a new technique for features extraction with adaptive evolving clustering method (ECM) of e-mails to build the evolving rules.
3. Minimize the complexity of the rule-base and configuration based on decrease the size of feature vectors from 21as "long vector" to four feature groups, which we called "short vector".
4. Achieves high performance, including high level of true positive, true negative, recall, precision, f-Measure and overall accuracy. Results improvement between 3% and 13% compare of the generated existing solutions to zero-day phishing e-mail exploits. The result was published in (ALmomani 2012; Almomani, Wan et al. 2012).
5. Decrease the average time consumed and the fuzzy rule generated by the short vector were about 10 times less than long vector.

## 7. Scope

The scope of this work is limited to detect `phishing emails` from ham emails in an online mode. The proposed framework sets between Message Transfer Agent (*MTA*) and Mail User Agent (*MUA)* to stop phishing email from reachable to victim account. *MTA:* acts the post offices and it represent the storing area and email carrier, message delivery agents (*MDA)*: act as mailboxes, which store messages (as much as their volume will allow) until the recipients check the box. *MUA*: a software program using for retrieving email like "Microsoft outlook". Figure 1.1 handles phishing message transportation. While *Phisher*: member sends a phishing message to a potential victim, *Victim*: the user may open the phishing email.

**Fig.1.** *Phishing Email Scope*

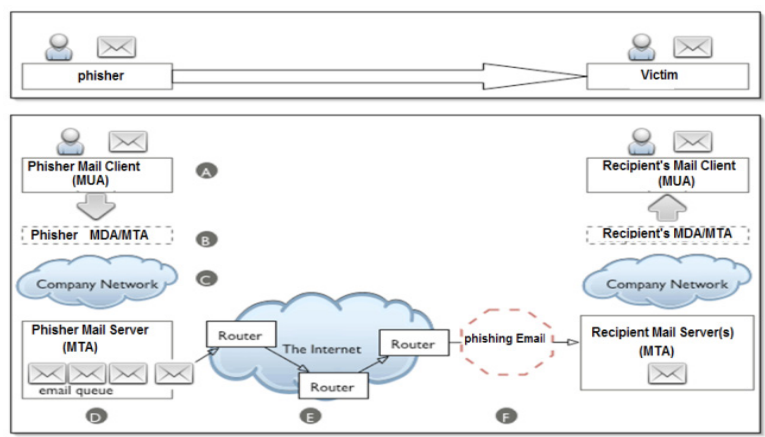

## 8. proposed Framework (PDENF)

Our proposed suggest a novel idea related to new framework called Phishing Dynamic Evolving Neural Fuzzy framework (PDENF), which able to detect and predict ("zero-day") phishing email in online mode with real implementation based on evolving connectionist system (ECoS)(Kasabov 2003),

ECoS is a connectionist architecture try to make easy of evolving processes with knowledge discovery. It can be neural network or set of networks, work continuously in time and adapt their structure and functionality through continuous relations with the environment and other systems. We proposed a hybrid (supervised /unsupervised) learning approach take the advantage of machine learning and fuzzy logic, with consideration the level of similarity between features of phishing emails (Almomani 2012).

Our proposed framework is shown in Fig. 2. In the proposed framework, ECOS is adapted based on the level of similarity among the four groups of phishing e-mail features. The proposed methodology is divided into four stages. The first stage is called pre-processing, used to extract 21 binary features from e-mails, which we called "*long vector*". The second stage is the e-mail object similarities used to decrease the size of feature vectors from 21 to four feature groups, which we called "*short vector*". The third stage includes the ECM and its offline extension (ECMc) to generate the basis of rules(Song and Kasabov 2001; ALmomani 2011). Finally, DENFIS is utilized in online mode as a fuzzy inferences system to create, update, or delete a fuzzy rule while the system is running. DyNFIS is also used in offline mode to enhance the rules in offline mode, enhance the level of classification accuracy, and decrease the error rate in the prediction process based on Gaussian membership function(Hwang and Song 2009). However, the profile management framework is suggested to put order the relationship between DENFIS and DyNFIS and employ the best rules in our framework.

The most effective 21 extracted features generated by many authors were adopted included feature in (Toolan and Carthy 2010; Khonji, Jones et al. 2011),some of the sub-features will merge into one feature, while the groups of features included Spam features, Body-based features , URL based Features and Features Header. The framework suggested working in Life-long working explained clearly in figure 3.

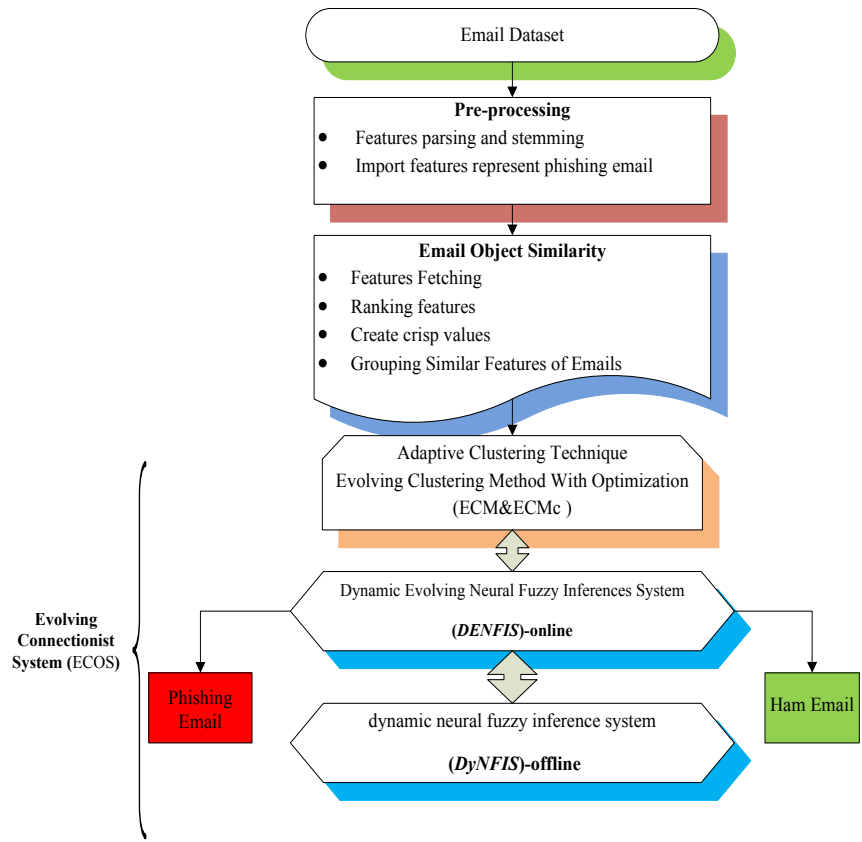

**Fig.2.** *PDENF*

**Fig.3.** *PDENF workflow scheme in clustering email server- Life-long working*

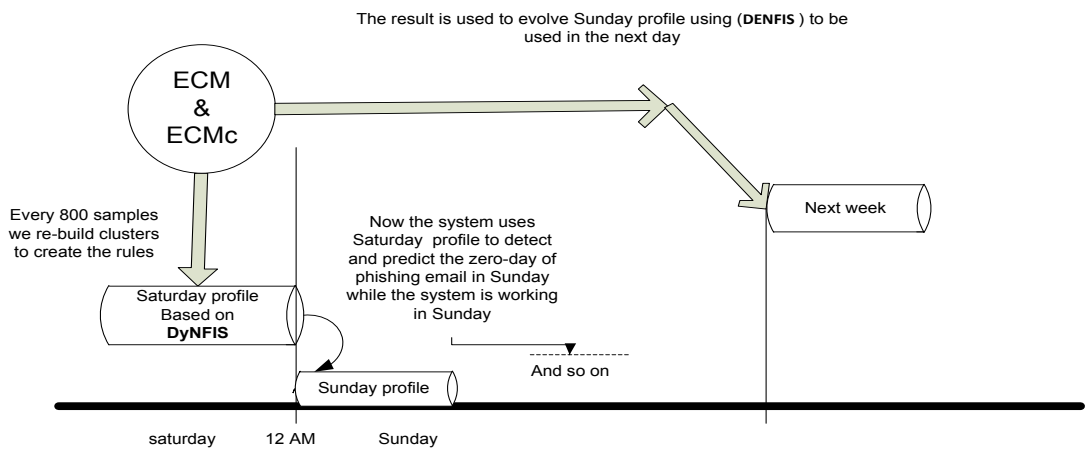

Figure3. Explains how the proposed framework suggested to be working as life-long working with footprint consuming memory and time. To do that we need system learn continuously in online mode based on *DENFIS* with enhance the rule after capturing pro-file consist in each time of 800 samples of email in offline mode based on *DyNFIS*. The profile management framework will load the enhanced rule again to DENFIS while the system is working. DENFIS will use the new enhanced rules automatically because it has the same format of rules, which depend on the Gaussian membership function. This will continue daily to load the full rule in Saturday to be used in Sunday while the system in working in unlimited way.

## 9. Conclusion

This proposal proposes a new framework called Phishing Dynamic Evolving Neural Fuzzy Framework (PDENF). Our framework expect to detect and predict unknown" Zero days" Phishing Email with decrease the level of false positive rate of ham email and false negative rate of phishing emails. This is to increase the level of accuracy and increase the performance of classification and prediction the phishing email values in online mode, and long-life working with footprint consuming memory.

## 10. Acknowledgments

This research is supported by National Advanced IPv6 Centre of Excellence (NAV6), Universiti Sains Malaysia (USM). Grant title: "A comprehensive botnet mitigation Ecosystem". Acc.No:1001/PNAV/857001.